\documentclass[fleqn,twoside]{article}
\usepackage{espcrc2}

\usepackage{graphicx}
\usepackage[figuresright]{rotating}

\newcommand{\AmS}{{\protect\the\textfont2
  A\kern-.1667em\lower.5ex\hbox{M}\kern-.125emS}}

\title{New advances in numerical simulations of $\theta$-vacuum systems}

\author{
V. Azcoiti \address[zara]{Departamento de F\'{\i}sica Te\'orica, 
Universidad de Zaragoza, E-50009 Zaragoza, Spain},
G. Di Carlo\address[lngs]{INFN, Laboratori Nazionali del Gran Sasso, 
67010 Assergi, (L'Aquila) Italy},
A. Galante\addressmark[lngs]\address[aq]{Dipartimento di Fisica 
dell'Universit\`a di L'Aquila, 67100 L'Aquila, Italy}
        \thanks{Talk presented by A. Galante.},
V. Laliena\addressmark[zara].
}

\begin{document}

\begin{abstract}
We proposed two different methods for simulating theta-vacuum systems.
Both can be affected by systematic effects but these errors can be 
controlled by comparing the results of the two unrelated methods.
This has been done for some analytically soluble models and here
the same procedure is successfully applied to the interesting case of
$\textrm{CP}^9$ in the scaling region.
\vspace{1pc}
\end{abstract}

\maketitle

Quantum Field Theory with topological terms is a special case of
complex Euclidean actions: the action is naturally divided in a real
part and a pure imaginary one, the latter in the form of $i$ times
$\theta$ times an integer that depends on the gauge configuration.
The same structure is shared by spin systems coupled to imaginary
magnetic field: they do not describe physical systems but are 
nevertheless useful to improve our understanding of complex actions.

For this class of models we proposed originally a method for 
reconstructing the $\theta$ dependence from real action simulations.
The main ingredient was to compute the probability distribution
function of the topological charge at $\theta=0$ from simulations
at imaginary $\theta=ih$ \cite{prl}. 
An analytical form of the p.d.f. is extracted 
from the measurement of the mean value of the density of topological 
charge as a function of $h$. This is done using the saddle point 
approximation and doing a fit of numerical data with a suitably
chosen function. At the end a multi-precision code is used to
compute the sum over the different topological sectors and calculate
the relevant physical quantities.
The arbitrariness of the fitting function as well as
the use of saddle point relation (valid up to ${\cal O}(1/V)$ corrections 
at finite volume) is a possible source of uncontrolled systematic effects.
This is not the case in some soluble models considered (2D compact
$U(1)$ at any coupling as well as 1D AF Ising models coupled to
imaginary magnetic field) where the results where reproduced at $\%$ level
(or better) but in general it is not possible to put any upper bound on the 
magnitude of these systematic effects.

This was precisely our motivation to construct an alternative and 
independent approach \cite{plb}. 
This second approach requires as input the same
data as the first one $i.e.$ $x(h)$, the measurement of the mean value of 
the density of topological charge as a function of $h$. 
This time we do not try to reconstruct the p.d.f. but use the fact that
the sum over topological sectors that defines the partition function
can naturally be written as an even polynomial in the variable
$z=\cos\frac{\theta}{2}$ and the derivative of the free energy density 
respect to $\log z^2$ gives $y(z)=x(\theta)/\tan\frac{\theta}{2}$
where $x(\theta)$ is the topological charge density as function 
of $\theta$.
From the last relation and its analytical continuation to imaginary 
$\theta$ we see that the behavior of $y(z)$ for 
$z=\cosh(h/2)\in [1,\infty)$ is related to $x(h)$ at $h\ge 0$ while for 
$z=\cos(\theta/2)\in [0,1]$ we can get  $x(\theta)$.
From real action simulations we can get $y(z)$
for any value of $z\ge 1$, $i.e.$ up to
$y(z=1)=2\chi_o=2\frac{dx(\theta)}{d\theta}|_{\theta=0}$
and since 
$y(z\to 0)=\lim_{\theta\to\pi}\frac{x(\theta)}{\tan(\theta/2)}$ 
the behavior of $y(z)$ near the origin is related to $x(\theta\simeq\pi)$.
Using the transformation $y_\lambda(z)=y(e^{\lambda/2}z)$ we can define
an effective exponent 
$\gamma_\lambda=\frac{2}{\lambda}\log \left( \frac{y_\lambda}{y}\right)$
that can help to distinguish between different behaviors at 
$\theta\sim\pi$ since if $q(\theta\to\pi)\propto (\pi-\theta)^\alpha$
then $\gamma_\lambda(y\to 0)=1+\alpha$. 
If $\gamma_\lambda(y\to 0)=1$ $\textrm{CP}$ symmetry is spontaneously 
broken at $\theta=\pi$, values between 1 and 2 indicate a CP symmetric 
vacuum with a divergent susceptibility (second order phase transition)
and $\gamma_\lambda(0)=2$ analiticity of the free energy at $\theta=\pi$.
It is possible to measure $\gamma_\lambda$ only for values of $y$ larger
than $2\chi_o$ and our experience with several models is that this is
a smooth function of the argument. 

Our second method is based on the
feasibility of a simple extrapolation procedure inside the 
{\it forbidden region} to reconstruct the full $\gamma_\lambda$ 
dependence. Once we have the effective exponent at any $y>0$ we
can reconstruct the topological charge density dependence on $\theta$
using an iterative procedure as explained in \cite{plb}.

\begin{figure}[t]
\vspace{9pt}
\centerline{\includegraphics*[width=2.3in,angle=90]{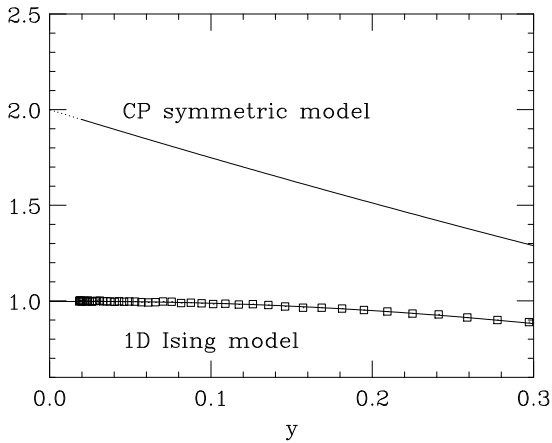}}
\caption{The effective exponent $\gamma_\lambda$. 
Lower curve: the symbols are from simulation and the continuous line is 
the quadratic fit of data.
Upper curve: the continuous line is the analytic result for imaginary
$\theta$ and the dotted part corresponds to real $\theta$ solution.}
\label{fig1}
\end{figure}

As an example we plot in Fig. 1 the effective exponent for 1D AF Ising
model ($\textrm{CP}$ symmetry is spontaneously broken at $\theta=\pi$)
and for a toy model that resembles the free instanton gas solution
(the partition function is ${\cal Z}_V(\theta)=(1+A \cos\theta)^V$
where $A$ is a constant and the $\textrm{CP}$ symmetry is enforced at  
$\theta=\pi$). In both cases a simple quadratic fit allows to extract
values of $\gamma_\lambda(y=0)$ compatible
with 1 and 2 respectively (with error at $.1\%$ level). 
Also the reconstruction of the order parameter
gives results with errors at level of few $\%$ maximum (see \cite{plb}). 

This shows that the extrapolation of the effective exponent is a
reasonable possibility. Of course in doing that we assume a smooth
behavior near the origin and this is equivalent to assume that no
phase transitions are present as a function of $\theta$ except at
most at $\theta=\pi$. We can also naively expect that the extrapolation
is safer for asymptotically free models since in this case the numeric
data can extend to smaller values of $y$ since the topological charge
susceptibility $\chi_o$ goes to zero increasing $\beta$.
Since $\lambda$ is a free parameter in the game an internal consistency 
check of the procedure is to verify the independence of physical 
observables on it.

Clearly, once again, systematic effects introduced by the arbitrary
choice of the extrapolating function can be present.
What is important is that we expect the systematics to be different
for the two different methods. 
On the other hand, if both methods give consistent results,
our confidence on the quality of the results is very high.

Once we have two methods to crosscheck our results the main goal
would clearly be to understand the role of the $\theta$ parameter
in QCD and its connection with the strong $\textrm{CP}$ problem.
The application of our methods to QCD with theta-term would
be straightforward provided we have a practical definition of the   
topological charge in order to do simulations at imaginary $\theta$. 
It turns out not to be the case: the known 
regularizations of the topological charge (e.g. ginsparg-wilson) are so 
involved that at the moment they are not of practical use.
A good possibility for studying the $\theta$-term effect on the low 
energy dynamics are two dimensional $\textrm{CP}^{N-1}$ models since
they have $\theta$-vacua, a well defined topological charge 
and many features in common with four-dimensional Yang-Mills theories:
they are asymptotically free, possess instanton solutions, and 
admit a $1/N$ expansion.

We already considered $\textrm{CP}^{3}$ in the strong coupling region
and are now interested to extend these studies to the continuum limit.
In this case the effect of dislocations has to be 
negligible not to spoil the scaling properties of the theory.
From $\theta=0$ simulations we know that $N=10$ is large enough
\cite{campostrini}
thus we concentrated our efforts on the scaling region of the 
$\textrm{CP}^9$ model.

\begin{figure}[t]
\vspace{9pt}
\centerline{\includegraphics*[width=2.3in,angle=90]{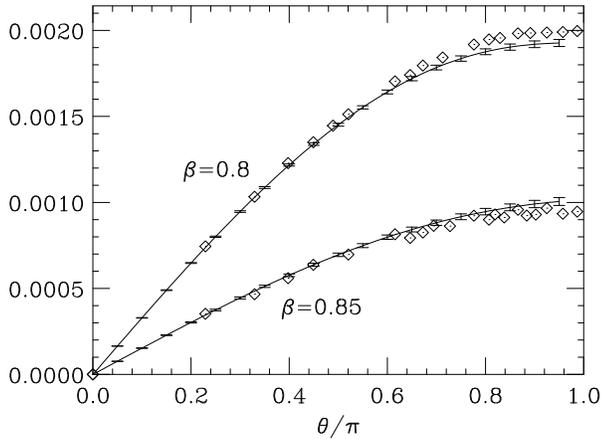}}
\caption{The topological charge density in $L=200$ lattice.
The points refer to the first method, the diamonds to the second one
(statistical errors are smaller than the symbol).}
\label{fig2}
\end{figure}
We have adopted for
the action the ``auxiliary U(1) field'' formulation
and the topological charge operator is defined directly from the $U(1)$ field
(see \cite{plb}).
For $N=10$ and $h=0$ the scaling window starts at $\beta$ around 0.75 
and we decided to concentrate our efforts at $\beta=$0.8 , 0.85 , 0.90.
The lattice sizes considered are
$L=100, 200$ for the two larger couplings and $L=200$ for the smaller one.
A standard Metropolis algorithm was used to generate ${\cal O}(10^6)$
configurations for each value of $\beta$ and $h$ (we
refer to \cite{plb} for a full discussion of those details).

In Fig. 2 we compare the results of the two methods for two values 
of $\beta$: the agreement is at (few) $\%$ level, compatible with
the statistical one (the errors in the plot are the statistical ones).
Following our previous consideration this can be considered the
strongest possible {\it a posteriori} consistency check of the results.

We used our high quality data to address the scaling properties of the
theory too. This has been done perturbatively and non 
perturbatively \cite{cp9}.
In Fig. 3 we show how the non perturbative scaling (only data from the 
first method are plotted) is realized up to violations that are at most
$\pm 4\%$. As expected the biggest violations are for the smaller
inverse coupling. We can conclude that we have strong indications
that, for this model, $\textrm{CP}$ is spontaneously broken in the 
continuum, as predicted by the large $N$ expansion.
\begin{figure}[t]
\vspace{9pt}
\centerline{\includegraphics*[width=2.3in,angle=90]{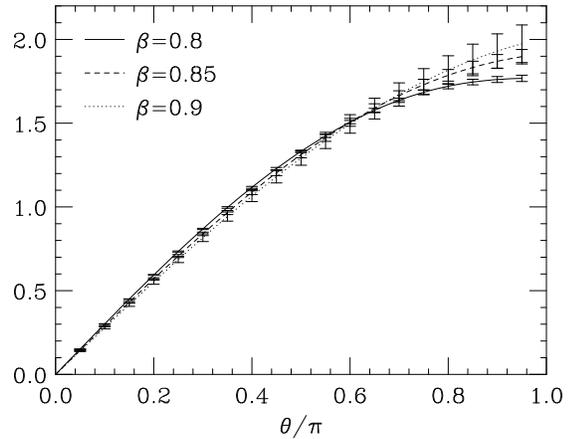}}
\caption{The topological charge density divided by the measured
topological susceptibility at $\theta=0$ for the $L=200$ lattice.}
\label{fig3}
\end{figure}

\end{document}